\documentclass[%
 reprint,
nofootinbib,
 amsmath,amssymb,
 aps,
]{revtex4-1}
\usepackage{float}
\usepackage{graphics}
\usepackage{graphicx}
\usepackage{dcolumn}
\usepackage{bm}

\begin{document}

\preprint{APS/123-QED}

\title{Rational function regression method for numerical analytic continuation}

\author{Jian Wang}
\author{Sudip Chakravarty}%
\affiliation{%
 Department of Physics and Astronomy, University of California, Los Angeles, California 90095\\
 Mani L Bhaumik Institute for Theoretical Physics 
}%

\date{\today}

\begin{abstract}
A simple method for numerical analytic continuation is developed. It is designed to analytically continue the imaginary time (Matsubara frequency) quantum Monte Carlo simulation results to the real time (real frequency) domain. Such a method is based on the Pad\'e approximation. We modify it to be a linear regression problem, and then use bootstrapping statistics to get the averaged result and estimate the error. Unlike maximum entropy method, no prior information is needed. 
Test-cases have shown that the spectrum is recovered for inputs with relative error as high as 1\%.

\end{abstract}

\maketitle


\section{Introduction}

One of the bottlenecks of quantum Monte Carlo study is how to perform a reliable analytic continuation from imaginary time to real time. Two major families are the  Pad\'e method~\cite{Vidberg1977,PhysRevB.61.5147,0256-307X-34-7-077102,PhysRevB.86.235107,PhysRevB.93.075104,PhysRevLett.75.501} and the  kernel based maximum entropy method~\cite{JARRELL1996133,PhysRevB.41.2380,PhysRevE.81.056701,LEVY2017149,1806.03841,2017InvPr..33k5007A,1810.00913,PhysRevE.95.061302,PhysRevB.95.014102,PhysRevB.57.10287,cond-mat/0403055,PhysRevE.94.063308,PhysRevB.78.174429}. They both have their pros and cons \cite{PhysRevB.82.165125,PhysRevB.94.245140}. The Pad\'e method needs very accurate imaginary time input data.  The maximum entropy method requires {\em a priori}  information.

In this work, we are going to use the Pad\'e method by casting it to a standard rational function regression problem. In order to estimate the error, bootstrapping statistics is used to generate an ensemble of imaginary input data. Compared with the traditional kernel based method, the rational function representation is more natural, because the zeros and poles will capture  all the information, if the physical system is made of finite elements of RLC (resistor-inductor-capacitor) components.

\section{The statement of the problem}

\subsection{Notations}
The spectral function $\rho(\omega)$ is a $\mathbb{R}\rightarrow\mathbb{R}_{\geqslant 0}$ function. The analytic Green function $G(z)$ is a $\mathbb{C}\rightarrow\mathbb{C}$ function. They are related by:
\begin{eqnarray}
G(z)=\int_{-\infty}^{+\infty} \rho(\omega) \frac{1}{\omega-z} d\omega
\label{eq:Cauchy_integral}
\end{eqnarray}
The analytic Green function, is compact and elegant, because the Matsubara Green function and the retarded Green function can be represented as imaginary and real  part of the analytic Green function:
\begin{eqnarray}
G^M(\omega_n)=G(i\omega_n) \label{eq:Matsubara_Green}\\
G^R(\omega)=G(\omega+i0^+) 
\end{eqnarray}
The spectral function
\begin{eqnarray}
\rho(\omega)=-\frac{1}{\pi} \text{Im}[G^R(\omega)]
\end{eqnarray}
contains all the information of the dynamics. And it is easy to get the entire $G(z)$ from $\rho(\omega)$  by integrating Eq.~\ref{eq:Cauchy_integral} directly.
However, it is hard to recover $\rho(\omega)$ from the information of $G^M(\omega_n)$ via:
\begin{eqnarray}
G^M(\omega_n)=\int_{-\infty}^{+\infty} \rho(\omega') \frac{1}{\omega'-i\omega_n} d\omega'
\end{eqnarray}
This is an inverse problem.
$G^M$ comes from Monte Carlo simulation, with error,
and $\omega_n$'s  are discrete and finite.

\subsection{\label{chap:statement}Statement}
\textbf{Input:}
estimated Matsubara frequency Green function with the error $G^M(\omega_n)\pm\delta G^M(\omega_n) $ for  $ \omega_n \in      \{  \Omega, 2\Omega, \cdots, N\Omega \}$, where $\Omega=\frac{2\pi}{\beta}$ \footnote{We are using Boson Matsubara frequencies  throughout this paper. However, identical considerations apply for  Fermionic Matsubara frequencies.}

\textbf{Output:}
the estimated spectral function and its uncertainty $ \rho (\omega)\pm \delta\rho(\omega)$

In a better treatment, the error of Matsubara Green function would be an $N\times N $ co-variant matrix. But here we treat $G^M(\omega_n)$ as independent random variables.

\subsection{Test}
A good method  to test is as follows:

\textbf{Generation:} choose a test function $\rho_{\text{True}}(\omega)$

\textbf{Encryption:} use Eq.~\ref{eq:Cauchy_integral} and \ref{eq:Matsubara_Green} to generate $G^M_{\text{True}}(\omega_n)$, then add random noise $\delta G^M(\omega_n)$ to get $G^M(\omega_n) $

\textbf{Recovery:} using $\delta G^M(\omega_n)$ and $G^M(\omega_n) $ in the last step as input,  use the Pad\'e Regression method to get the output $\rho(\omega)\pm \delta \rho(\omega)$

\textbf{Comparison:} compare the recovered $\rho(\omega)$ and the original $\rho_{\text{True}}(\omega)$

\section{The method}

\subsection{Rational function method}
The Pad\'e method assumes that the analytic Green function $G(z)$ takes the  form of a rational function
\begin{eqnarray}
	Q_{L,M}(z)=\frac{p_{L}(z)}{p_{M}(z)}=\frac{a_0+a_1 z + a_2 z^2 + ... a_{L} z^{L}}{b_0+b_1 z + b_2 z^2 + ... b_{M} z^{M}}
    \label{eq:Pade}
\end{eqnarray}
Where $L$ and $M$ are the degrees of the  polynomials; as a normalization convention, we shall also choose $b_0=1$.
The idea is to use $(L+M+1)$ complex parameters $\{a_0,a_1,\cdots,a_L,b_1,\cdots,b_M\}$ to represent an arbitrary $G(z)$. Instead of using the value of $\rho(\omega)$ on discrete $\omega$ to represent $G(z)$, as in the maximum entropy method.
An alternative form of Eq.~\ref{eq:Pade} can be more physicially meaninguful, it is given by:

\begin{eqnarray}
	Q_{L,M}(z)=\frac{p_{L}(z)}{p_{M}(z)}=\frac{a_0(z-A_1)\cdots (z-A_L)}{(z-B_1)\cdots (z-B_M)}
    \label{eq:Pade2}
\end{eqnarray}
There are $L$ zeros and $M$ poles, and a complex amplitude $a_0$. As a result of causality, $G(z)$ should be analytic in the upper half plane. In a reasonable regression, all $B_i$ should be in the lower half plane, or $B_i$ should be canceled by $A_j$ in the upper half plane.
Also, for physics problems with symmetry, the distribution of zeros and poles should  have those symmetries. This reduces the degrees of freedom of the parameters.

\subsection{The regression problem}
As a regression problem, our input data are $N$ Matsubara frequencies $z_n=i \Omega n$, and the values of Green function $ u_n = G(z_n)$ at these frequencies. The output are the coefficients  $a_i$ and $b_i$ in the rational polynomial of Eq.~\ref{eq:Pade}. There are \(N\) equations, and \( L+M+1 \) parameters to be fit.  \(N\) of those Eq.~\ref{eq:Pade} can be  written in a linear regression form: Eq.~\ref{eq:linearRegressionMatrix} and Eq.~\ref{eq:linearRegressionStandardForm} \footnote{the equal sign ``=" in Equation \ref{eq:linearRegressionMatrix} and  \ref{eq:linearRegressionStandardForm} should be understood in a linear regression manner: find \( \pmb{\beta}\), such that \( ||\pmb{X}\pmb{\beta}-\pmb{y} ||^2 \) is minimized }. Where the matrix \(\pmb{X}\) and the vector \(\pmb{y}\) contain input data, the vector \(\pmb{\beta} \) contains the parameters to be calculated. 
\begin{widetext}
\begin{eqnarray}
\label{eq:linearRegressionMatrix}
\left(
\begin{array}{ccc|cccc}
-u_1 z_1^1 & -u_1 z_1^2 & \hdots &  z_1^0  & z_1^1 & z_1^2 & \hdots \\
-u_2 z_2^1 & -u_2 z_2^2 & \hdots &  z_2^0  & z_2^1 & z_2^2 & \hdots \\
\vdots & \vdots &\vdots &\vdots & \vdots &\vdots & \vdots \\

\vdots & \vdots &\vdots &\vdots & \vdots &\vdots & \vdots \\
\vdots & \vdots &\vdots &\vdots & \vdots &\vdots & \vdots \\
\vdots & \vdots &\vdots &\vdots & \vdots &\vdots & \vdots \\
\vdots & \vdots &\vdots &\vdots & \vdots &\vdots & \vdots \\
-u_N z_N^1 & -u_N z_N^2 & \hdots &  z_N^0  & z_N^1 & z_N^2 & \hdots \\
\end{array} \right)
	\begin{pmatrix}
b_1 \\
b_2 \\
\vdots \\
\vdots \\
\hline
a_0 \\
a_1 \\
a_2 \\

\vdots

\end{pmatrix}=
	\begin{pmatrix}
		u_1 \\
		u_2 \\
		\vdots \\
		\vdots \\
		\vdots \\
		\vdots \\
		u_{N-1} \\
		u_N
	\end{pmatrix}
\end{eqnarray}
\end{widetext}
\begin{eqnarray}
\label{eq:linearRegressionStandardForm}
\pmb{X}_{N\times(L+M+1)}   \pmb{\beta}_{(L+M+1) \times 1 }= \pmb{y}_{N \times 1}
\end{eqnarray}
Eq.~\ref{eq:linearRegressionStandardForm} is the compact form of Eq.~\ref{eq:linearRegressionMatrix}

Sub-index  \(n\) in \(z_n\) and \(u_n\) labels one Matsubara frequency point, and they are all at the \(n\)-th row of \(\pmb{X} \).   \(  \big[ n^{\text{th}} \text{ row of }  \pmb{X} \big]\pmb{\beta}=u_n \)  has the same meaning of Eq.~\ref{eq:Pade}

\subsection{Choice of $L$ and $M$\label{Chap:Choice of L and M}}

Here $\pmb{X}$ is a vandermonte-like matrix, it is highly singular. As a rule of thumb, we choose:
$$
L\approx M \approx N/2
$$
The argument is as follows: For too large $M$ and $L$, the model might be over-fitting. In the case of $L+M+1=N$, the number of equations  is the same as the numbers of fitting parameters. For small $L$, and $M$, we are afraid that, there will not be enough poles and zeros to represent $G(z)$.
$L\approx M $ tends to cancel zeros and poles, and the real $G(z)$ only contain a few poles.
In a fully developed Bayesian method, both $\pmb{\beta}$ and $L,M$ are taken as estimation random variables. But for simplicity, we are going to choose the most representative $N/2$ value as $L$ and $M$

\subsection{Bootstrapping statistics}

In the  Pad\'e method \cite{Vidberg1977}, single  \(\pmb{X}\) and \( \pmb{y} \) are used to generate a single $\pmb{\beta}$ without error estimation. Here, we treat \(\pmb{X} \) and \( \pmb{y}\) as the mean value of a distribution with standard errors \(\delta \pmb{X}\) and \( \delta \pmb{y} \). These errors come from Monte Carlo result:   \(u_i\) and  \( \delta u_i \). 

The idea of bootstrapping statistics is to generate \textit{an ensemble} of input data: $ \{ \pmb{X} \} $ and $ \{ \pmb{y} \} $. Then perform the regression  individually  to get an ensemble of $ \{ \pmb{\beta}\}$, and then get a collection of spectrum $ \{ \rho(\omega) \} $. From the ensemble of  spectrum, we take the best estimation and its uncertainty as  $  \text{mean}\{ \rho(\omega) \} $ and $  \text{std}\{ \rho(\omega) \} $ (standard deviation).
Compared to the traditional model based regression and error estimation, bootstrapping is simple and natural --- various slightly different inputs are thrown into this black-box , then we check the difference among those output spectra. If those outputs are close to each other, it indicates the spectrum recovery is reliable.

Now, we need to generate those resamplings, $ \{ \pmb{X} \} $ and $ \{ \pmb{y} \} $. It is done by replacing the best values \(u_i\) in Equation (\ref{eq:linearRegressionMatrix}) by a distribution of themselves. Our assumption is that the Monte Carlo estimation of Green function value \(u_i\) has a normal distribution \( \mathcal{N}(u_i,\delta u_i) \). This is a result of the central limit theorem. In this paper, we take the relative error \(\frac{\delta u_i}{u_i} \leq 1 \% \). The assumption is even better satisfied for smaller relative errors. The procedure is summarized as follows:

\begin{itemize}
    \item Generate an ensemble of resampling $ \{\pmb{X}\}, \{\pmb{y}\} $.
    This is done by replacing the best value $u_i$  in Equation  (\ref{eq:linearRegressionMatrix}), with its distribution
   $ u_i \rightarrow \mathcal{N}(u_i,\delta u_i)= \mathcal{N}(\hat{G}^M(\omega_n),\delta_{G^M}(\omega_n) )$ 
   
    \item Perform least square linear regression for individual input data pair  \( \pmb{\beta}= \pmb{X}^{-1}\pmb{y}   \) 
    so that we have an ensemble $ \{ \pmb{\beta} \} $
    
    \item Use  $ \{ \pmb{\beta} \} $ to generate  $\{\rho(\omega)\}$ and then calculate $\text{mean}\{\rho(\omega)\} $ and $  \text{std}\{\rho(\omega)\} $
\end{itemize}

\begin{figure}[H]
	\centering\includegraphics[width=1.0\linewidth]{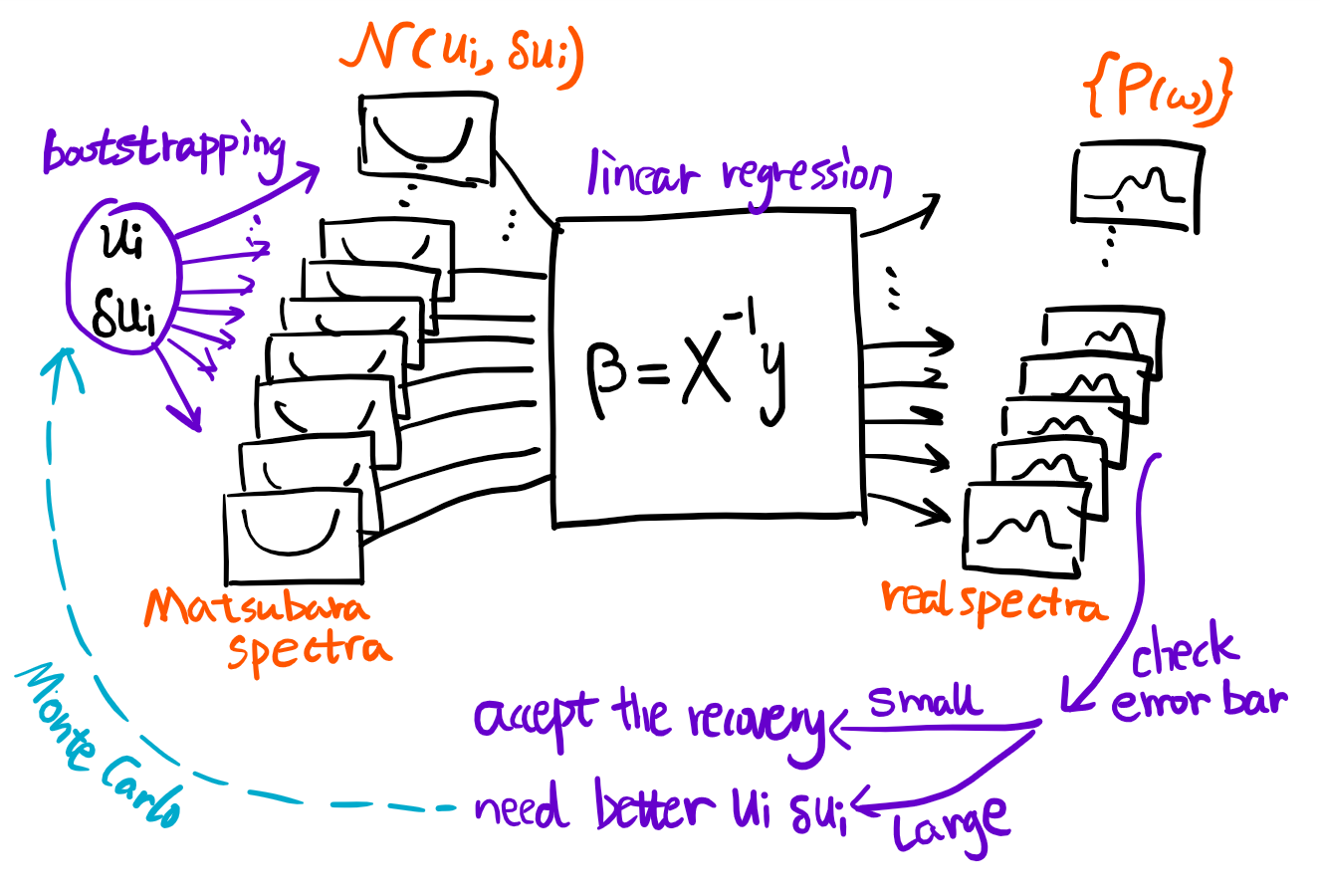} 
	\caption{  A schematic of the method. (1) The Monte Carlo data $u_i$ and $\delta u_i$ is used to generate an ensemble of Matsubara spectra. (2) Perform linear regression individually. (3) Get an ensemble of real spectra. Check the relative error of those spectra, if too large, then we need longer time Monte Carlo calculation for a smaller $ \delta u_i /u_i$. }\label{fig:spectrumRecovery}
\end{figure}

The number of resamplings is defined as $\mathfrak{N} $. It should be large enough, so that $\text{mean}\{\rho(\omega)\}$ and $\text{std}\{\rho(\omega)\}$ (std stands for standard deviation) converge. Our answer  is therefore given by
\begin{equation} \label{eq:answer}
\rho (\omega)\pm \delta\rho(\omega) \approx \text{mean}\{\rho(\omega)\} \pm \text{std}\{\rho(\omega)\}
\end{equation}
Notice that the estimated error $\delta\rho(\omega)  $ \textit{is not} $ \frac{\text{std}\{\rho(\omega)\} }{\sqrt{\mathfrak{N}}} $.  $\delta \rho(\omega)$ is the variation of the output spectrum, subject to slightly different input data. It represents the robustness of such ``input-blackbox-output" system (Fig.~\ref{fig:spectrumRecovery}), therefore should be $\text{std}\{\rho(\omega)\}$. Eq.~\ref{eq:answer} is  asymptotically reliable as the relative error $\delta \rho /\rho $ becomes smaller and smaller. If this relative output error is larger than  order 1  (for example $\frac{1}{3}$),  we need to continue Monte Carlo simulation for a higher precision \(u\pm \delta u \), and then use it as the input data of the blackbox.

Such bootstrapping doesn't take too much time to run, the major time cost still comes from Monte Carlo. In a problem with \(N=35\) Matsubara frequency points, taking \( \mathfrak{N}=20000 \) resamplings for good convergence, it only costs one minute in a laptop.  The overall time complexity is $O(\mathfrak{N} N^3 ) $, as it performs $O(N^3)$ linear regression   \( \pmb{\beta}= \pmb{X}^{-1}\pmb{y} \) for \( \mathfrak{N}\) times. The choice of discrete lattice $\omega$ only affects the plotting.

\section{Test cases}
Two factors can change the testing results, which  we should be aware of. The first factor is the number of Matusbara frequency data points \(N\) and the interval \( \Omega=\frac{1}{\beta} \).   They should be chosen such that the most of the spectral weight is within the range $ [-N\Omega , N \Omega ]$, or say $ \int_{+N\Omega}^{-N\Omega} \rho(\omega) d \omega \approx 1 (\text{normalized } \rho(\omega))$.  The second factor is the relative error of input data \( \eta = \delta u /u \). We take $\eta=10^{-15},10^{-6},10^{-4},10^{-3},10^{-2}$,   representing the error of most  diagrammatic expansion or  Monte Carlo simulation. In this section, we are going to use a piece-wise linear function (Fig~\ref{fig:piecewise0}) as the test spectrum; more test cases are given in  the appendix.
\subsection{Input data with small error}
\begin{figure}[H]
	\centering\includegraphics[width=\linewidth]{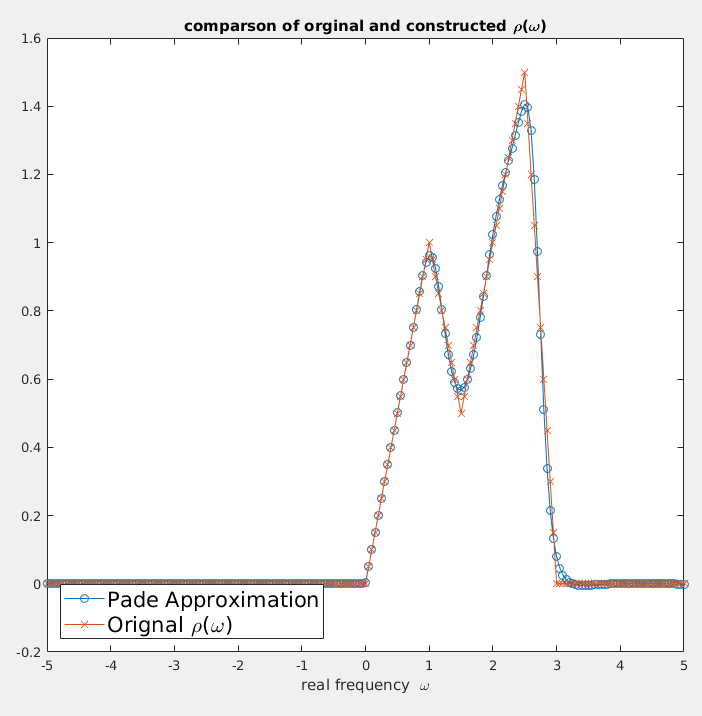}
	\caption{Pad\'e approximation spectrum recovery, number of input  Matsubara frequency points $N=35$;  error of input Matsubara frequency Green function $\eta=10^{-15}$. The orange curve is the piece-wise linear test function, the blue curve is the recovered spectrum. They agree very well, except for a few sharp tuning points }\label{fig:piecewise0}
\end{figure}

In Fig.~\ref{fig:piecewise0}, the orange curve is the exact test spectral function.  The blue curve is a Pad\'e recovery from blurred imaginary Green function with machine precision percentage error ($10^{-15}$). 

\begin{figure}[H]
	\centering\includegraphics[width=\linewidth]{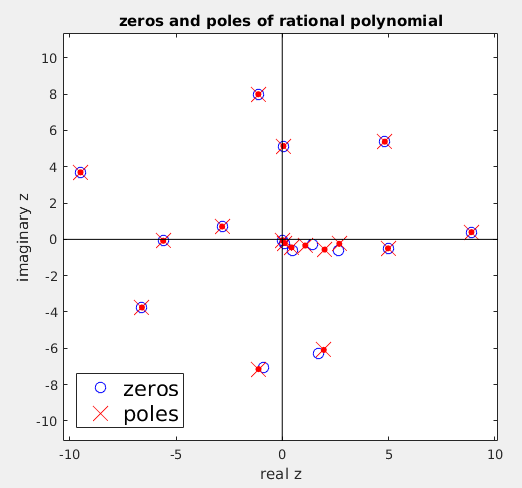}
	\caption{Distribution of zeros and poles. Input fitting points $N=35$ , degree of nominator polynomial $L=17$, degree of denominator polynomial $M=17$. There are exactly 17 poles on the complex plane. In the upper half plane, the zeros cancels the poles (removable singularity), thus making it analytic (causality). The poles close to the real axis has a huge influence on the shape of the approximation spectrum $\rho(\omega)$: they are the most important part of Pad\'e approximation. }\label{fig:zeropole}
\end{figure}

In Fig.~\ref{fig:zeropole}, a lot of poles and zeros  are paired together; it probably means that $N=35$ parameters correspond to over-fitting.  But this pairing-canceling mechanism makes the result robust, even for over-fitting parameters. This is also the reason why, we approximately chose  $L\approx M \approx N/2$   in section \ref{Chap:Choice of L and M}.  
Also, as a result of causality, the upper half plane should have no poles. We see that  all the poles are cancelled by zeros in the upper half plane. The locations of zeros and poles, and the coefficient  $a_0$ in Eq.~\ref{eq:Pade2} carry all the information. Actually, it is the zeros and poles, which are closest to the real axis that will mostly influence the shape of the spectral function. In other other words, if some zeros or poles are far away from the origin, it will have very little influence in the result. This is the second reason for the robustness. 

\subsection{Input data with large error\footnote{
$\eta=10^{-6},10^{-4},10^{-3},10^{-2}$ are large compared with $\eta=10^{-15}$}}

\begin{figure}[H]
 
 	\centering\includegraphics[width=1\linewidth]{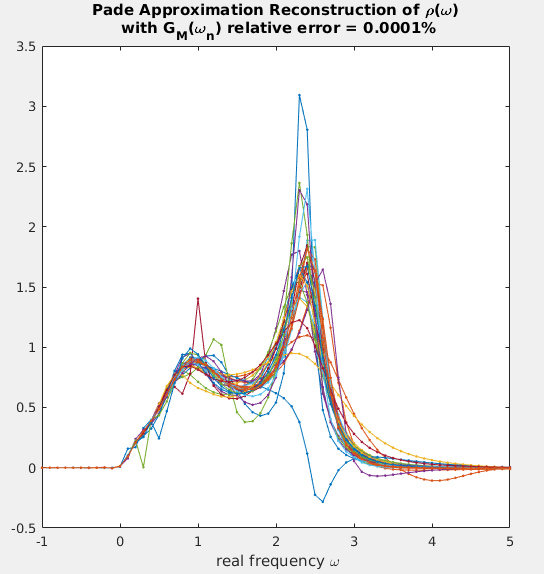}
 	\caption{The $\{\rho(\omega)\} $ ensemble. Each $\rho(\omega)$ curve is one recovery using Pad\'e regression, there are 30 of them in the plot. The input data's relative error is $\eta=0.0001 \%$ (6 significant digits). }\label{fig:noodles}
 \end{figure}

\begin{figure}[H]
	\centering\includegraphics[width=1\linewidth]{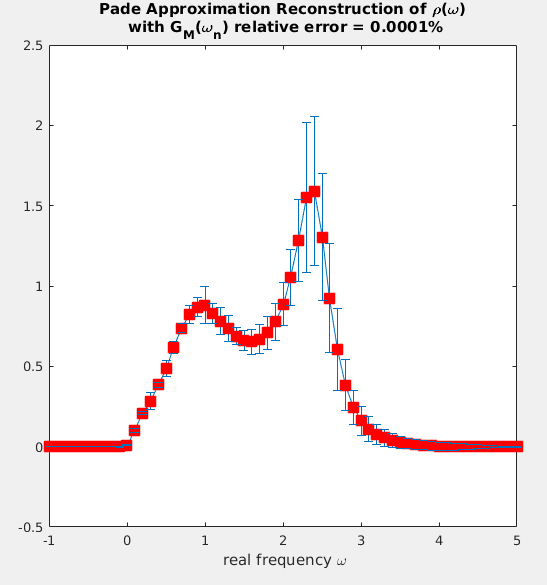}
	\caption{  6 significant digits input recovery.  Red dots are the $\text{mean}\{\rho(\omega)\}$, the blue error bars are  $\text{std}\{\rho(\omega)\}$ . The relative error of the output spectrum is less than $ \frac{1}{3}$, detailed shape is reliable.  }\label{fig:noodlesAverage}
\end{figure}

The computational time scales  as 
$\text{[CPU Time]}  \propto 1/{\sigma^2} = (\frac{u}{\delta u})^2 $. 
Clearly, we cannot have machine precision Monte Carlo data for really large systems.
Below is a test with large error in the input Matsubara frequency data.
Fig.~\ref{fig:noodles}  is an ensemble of recovered real frequency spectrum using bootstrapping statistics. The relative error of input Matsubara frequency data is $0.0001\%$. Fig \ref{fig:noodlesAverage} is the averaged value and error bars.

\begin{figure}[H]
	\centering\includegraphics[width=1\linewidth]{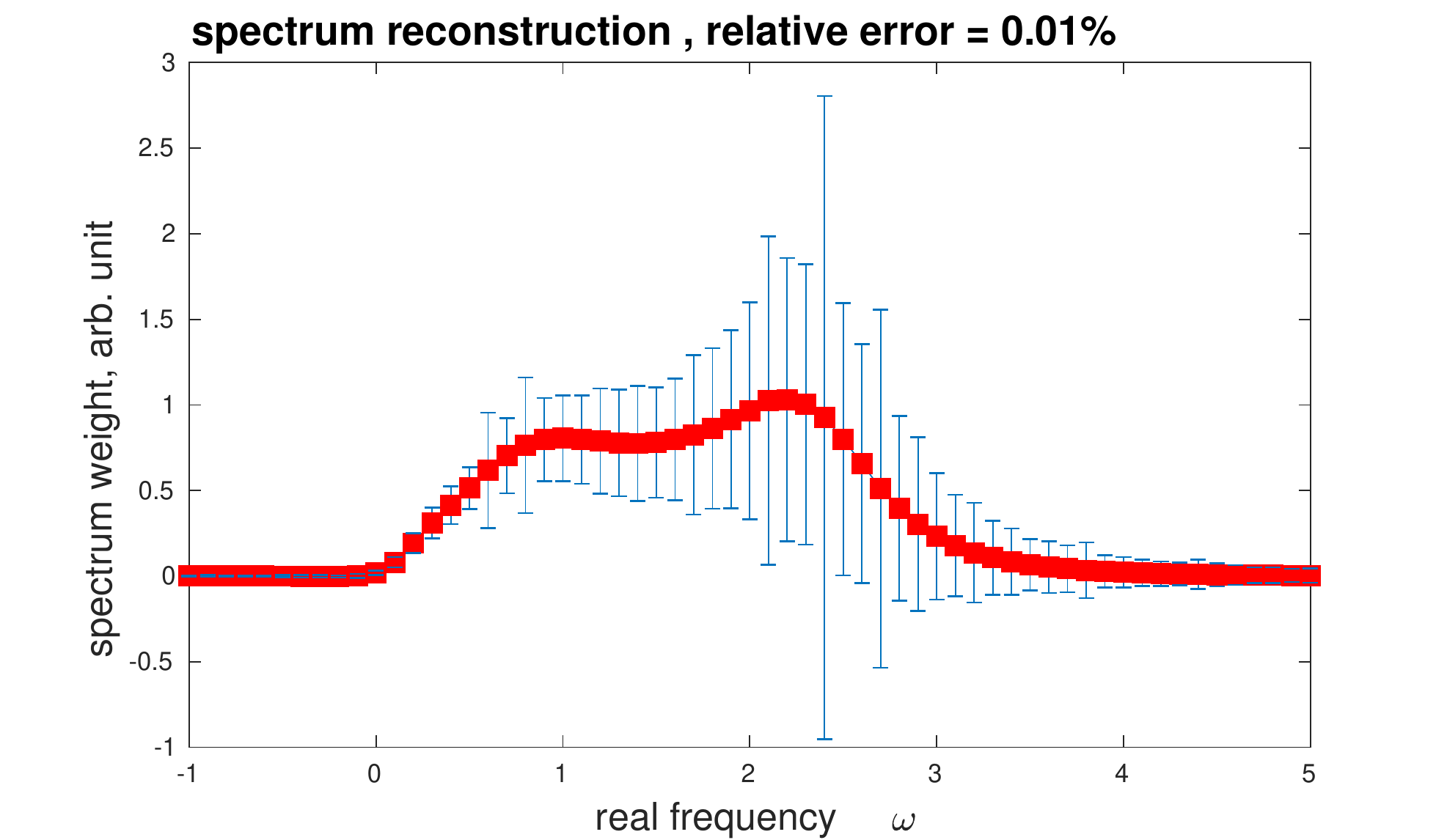}
	\caption{ 4 significant digits input recovery. Red dots are the $\text{mean}\{\rho(\omega)\}$, the blue error bars are  $\text{std}\{\rho(\omega)\}$ .  The relative error is of order one, the detailed shape is not reliable.
	} \label{fig:error4}
\end{figure}

\begin{figure}[ht]	
	\centering\includegraphics[width=1\linewidth]{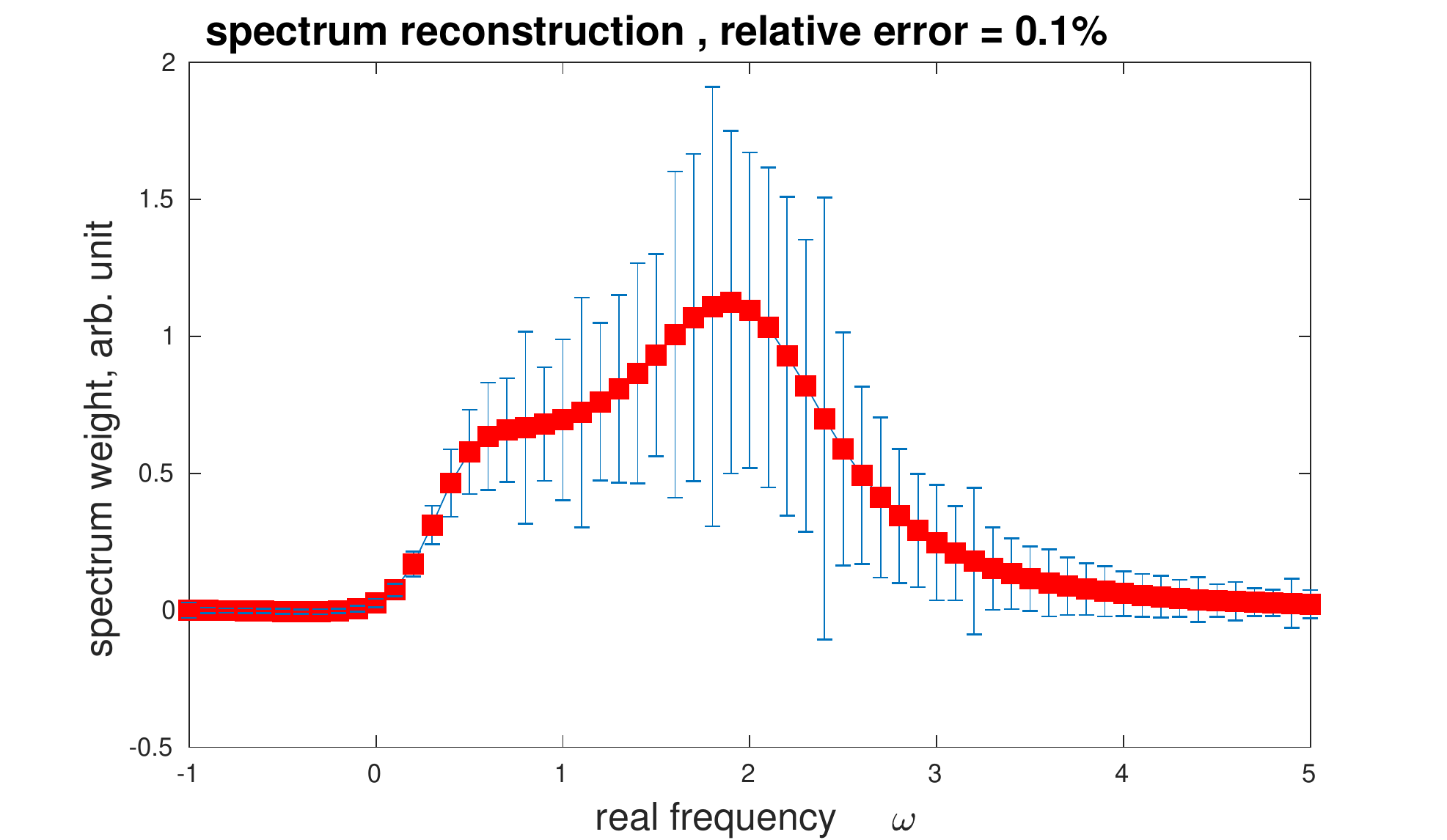}
	\caption{ 3 significant digits input recovery. Red dots are the $\text{mean}\{\rho(\omega)\}$, the blue error bars are  $\text{std}\{\rho(\omega)\}$. The relative error is of order one, the detailed shape is not reliable. } \label{fig:error3}
\end{figure}

\begin{figure}[ht]
		\centering\includegraphics[width=1\linewidth]{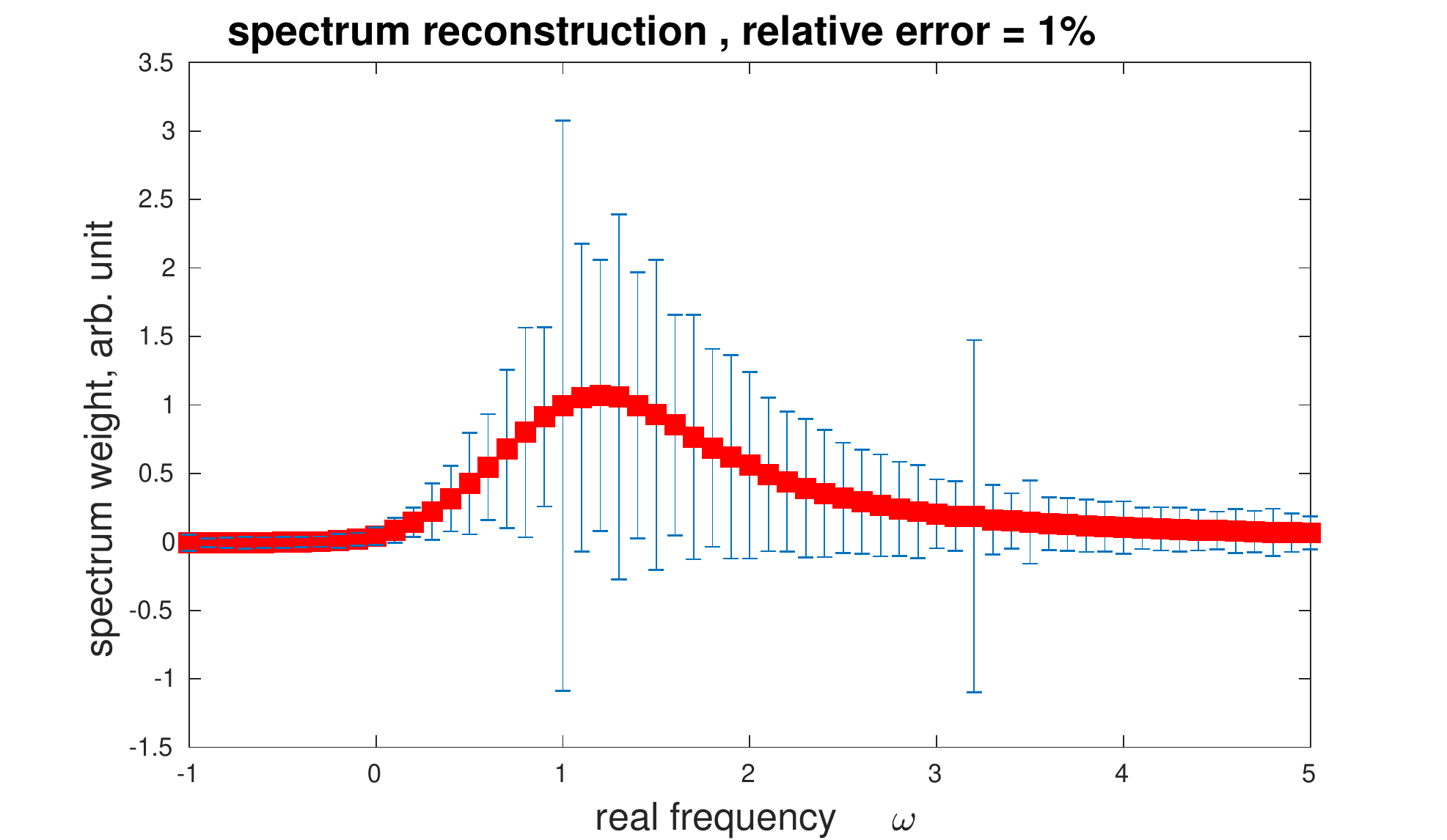}
	\caption{ 2 significant digits input recovery. Red dots are the $\text{mean}\{\rho(\omega)\}$, the blue error bars are  $\text{std}\{\rho(\omega)\}$. The relative error is at order one, the detailed shape is not reliable. } \label{fig:error2}
\end{figure}

Figures  \ref{fig:error4}, \ref{fig:error3}, and  \ref{fig:error2} give the results of relative error $ 0.01 \%, 0.1 \%, 1 \% $ respectively.
We can see that, the 0.01\% result still gives the accurate locations of double peaks \(\omega=1,2.5\), and the valley at \(\omega=1.5\), and linear shape of the curves. Even for the 1\% error data, our method generates a very reasonably recovered spectrum, it locates the spectrum's location \( 0< \omega <3 \) and gives the correct peak height around 1 to 1.5.  Notice that, for such test spectrum, double peak triangles, is a difficult function to recover.
In the appendix, a family of physically sensible spectrum are tested. 

\begin{figure}[ht]
	\centering\includegraphics[width=1\linewidth]{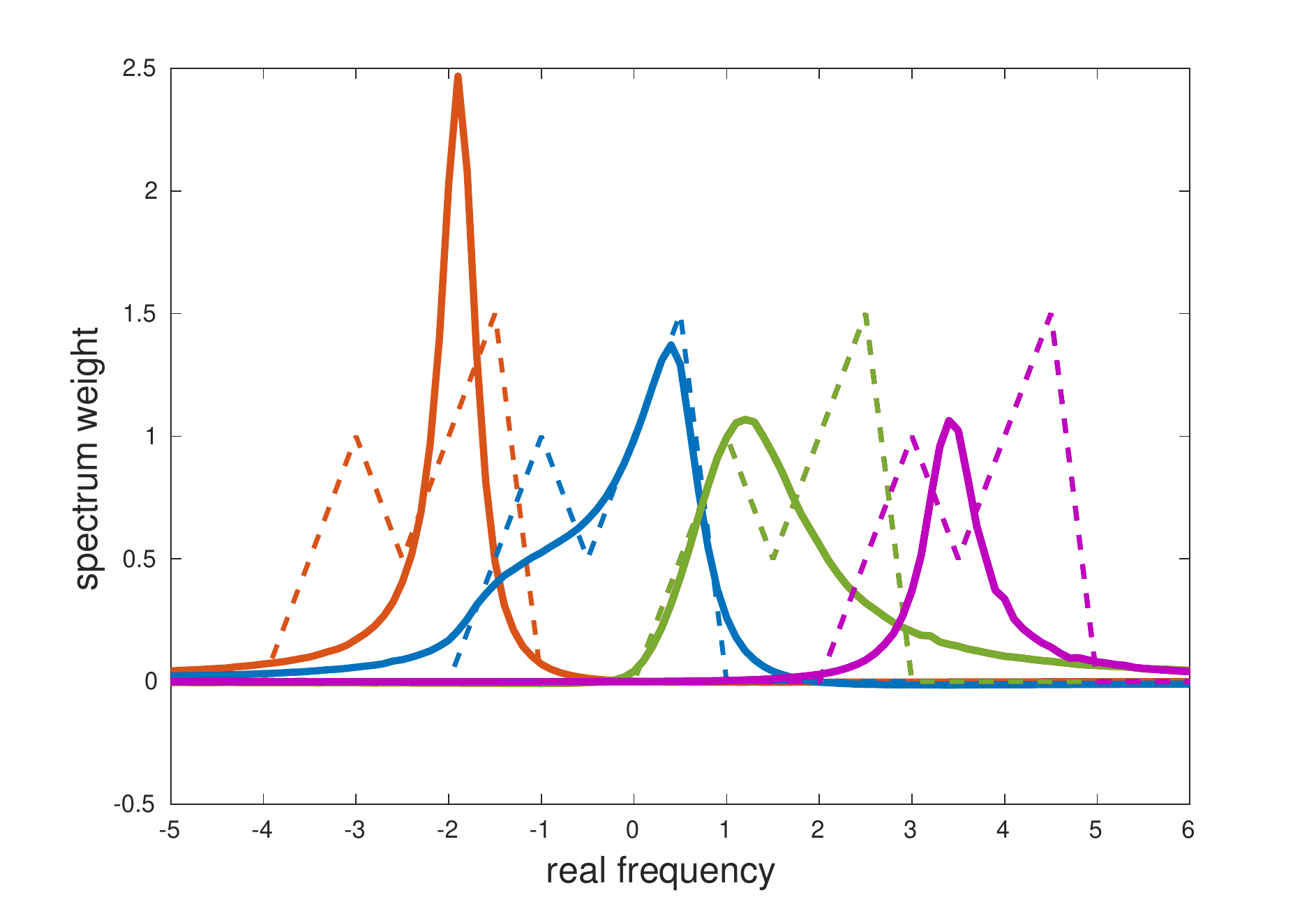}
	\caption{ Spectrum recovery for 1\% relative error input data.  Solid curves are the recovered spectrum; dashed curves are the original test function, they are shifted by 2 for comparison. There are 4 pairs, same color is the pair.  We can see that, the position of spectrum falls in the correct range of each original test function, the magnitude is also at the same order of the test function. The blue curve agrees reasonably well.  This method performs better for low frequencies, because low frequency points are closer to imaginary axis' input data. }\label{fig:1percent4figs}
	
\end{figure}

In order to check that $\eta=1\%$  recovery is not an accident,  we shift the double triangle spectrum horizontally by -4,-2,0,2 to get four difference test functions (Fig.~\ref{fig:1percent4figs}), we can see that the recovered spectrum all falls in the correct range. And the performance is surprisingly well for the lower frequency blue curve, because its spectral weight is closer to the imaginary axis.

However, if we want to recover the \textit{detailed shape} of an unknown spectrum, we should really check the error bar $\text{std}\{\rho(\omega)\}$. When the error bar is large (same order as the value), the detailed shape is not reliable, which is the case of Fig.~\ref{fig:error4}~\ref{fig:error3}~\ref{fig:error2}. In the case of Fig.~\ref{fig:noodlesAverage}, the error bar is no larger than $\frac{1}{3}$ of the best value, we are then sure that the detailed shape is reliable.
\section{Conclusion}
In this work we use rational function to represent the physical system. A matrix form is constructed, to convert it to a standard linear regression problem. Bootstrapping statistics is applied, to get best estimation and estimated errors.
For high precision recovery, the error gives information about whether or not we need to increase the Monte Carlo data's accuracy. For low precision recovery, our method still gives correct position and amplitude of the spectrum even for 1\% relative error input data.
This regression form can be used for further study, either combined with maximum entropy, or machine learning methods \cite{1806.03841,2017InvPr..33k5007A,1810.00913}. Future work can also be done utilizing the symmetry aspect of zeros and poles and the fully Bayesian choices of $L$ and $M$.
\begin{acknowledgments}
This work was supported by the funds from the David S. Saxon Presidential Chair. 
\end{acknowledgments}

\appendix

\section{Recovery test for more functions}
Below, a few other functions are given as examples. The dashed green thick line is the exact spectral function. The other 4 solid lines are Pad\'e regression recovered results for different relative errors, ranging from 1\% to 0.001\%
First of all, we see that, this method all gives the correct location of spectral weight, even for 1\% error.  Secondly, Lorentzian  curves are exactly recovered, (single peak 0.1 \%, double peak 0.001\%), because they are rational functions.
For the Gaussian curve, we cannot recover the detail shape, but the location of the peak is still accurate.
For the semicircle and square, the exact shapes are not recovered, but the starting and the ending frequencies agree reasonably well. As the error gets smaller, the more peaks is added to approach the exact result.

\begin{figure}[H]
	\centering\includegraphics[width=1\linewidth]{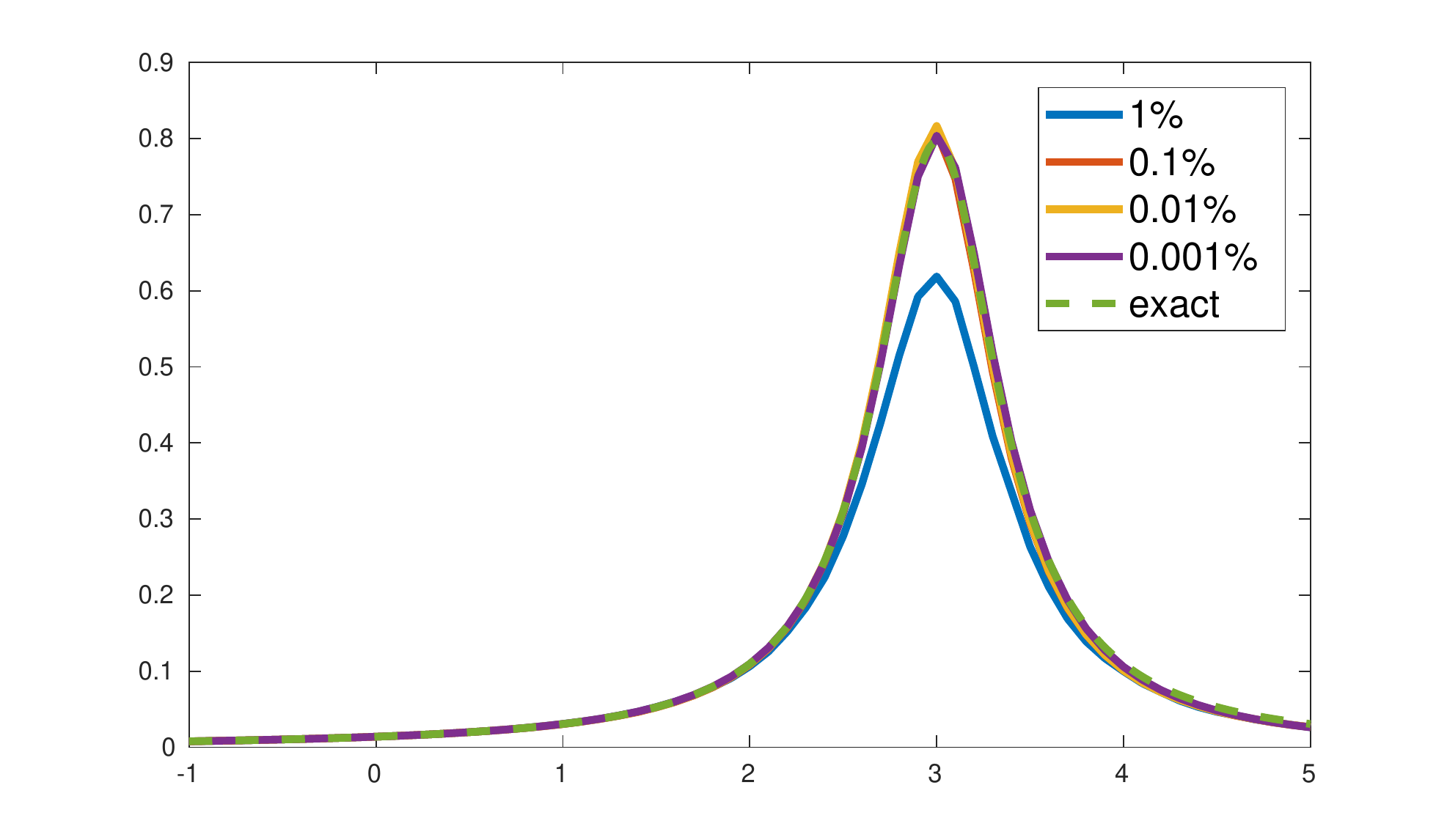}
	\caption{Dashed line is the Lorentzian distribution test spectrum. Colored solid lines are recovered spectra with different input errors.} 
\end{figure}

\begin{figure}[H]
	\centering\includegraphics[width=1\linewidth]{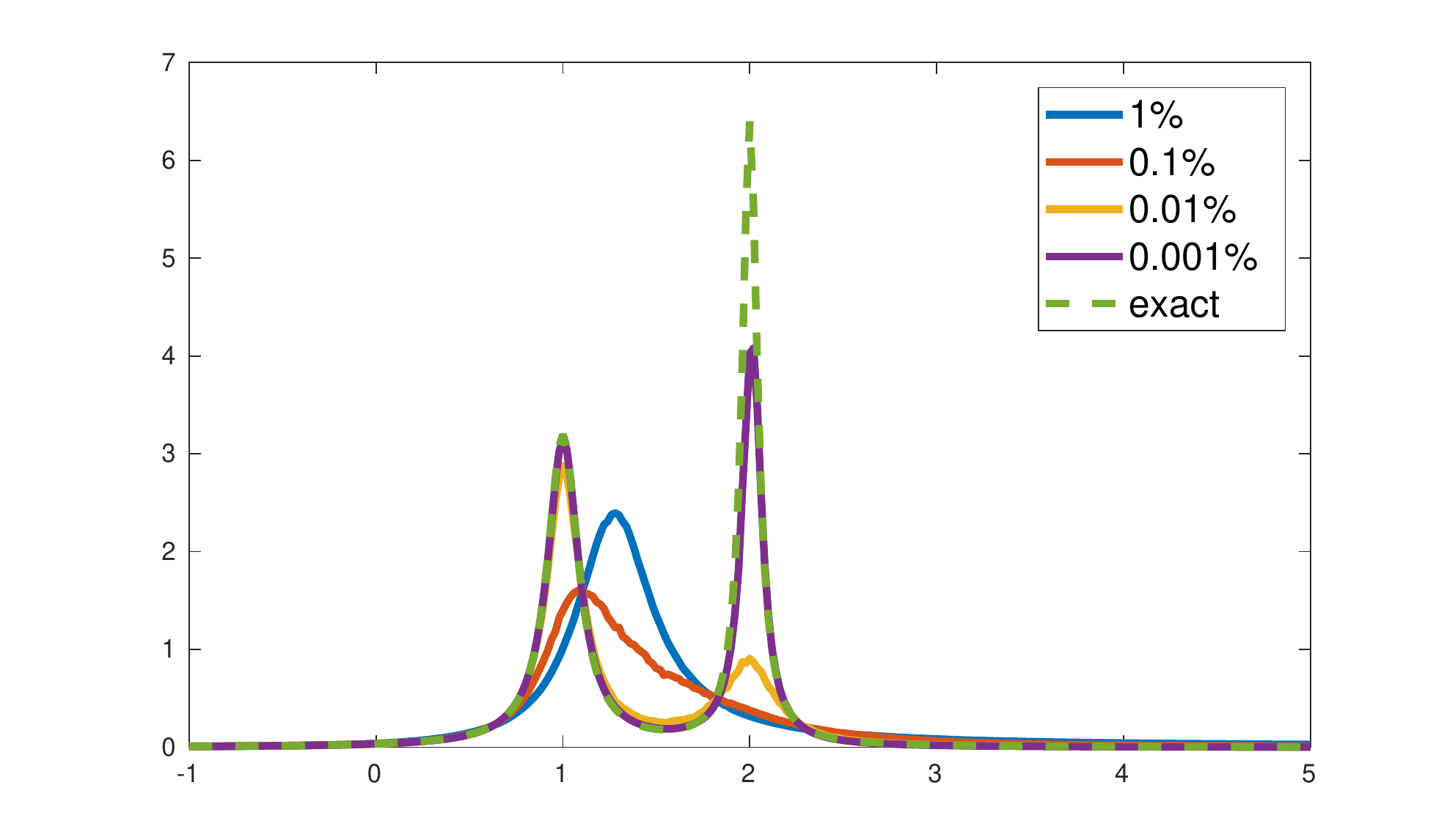}
	\caption{Dashed line is the double Lorentzian distribution test spectrum. Colored solid lines are recovered spectra with different input errors.} 
\end{figure}

\begin{figure}[H]
	\centering\includegraphics[width=1\linewidth]{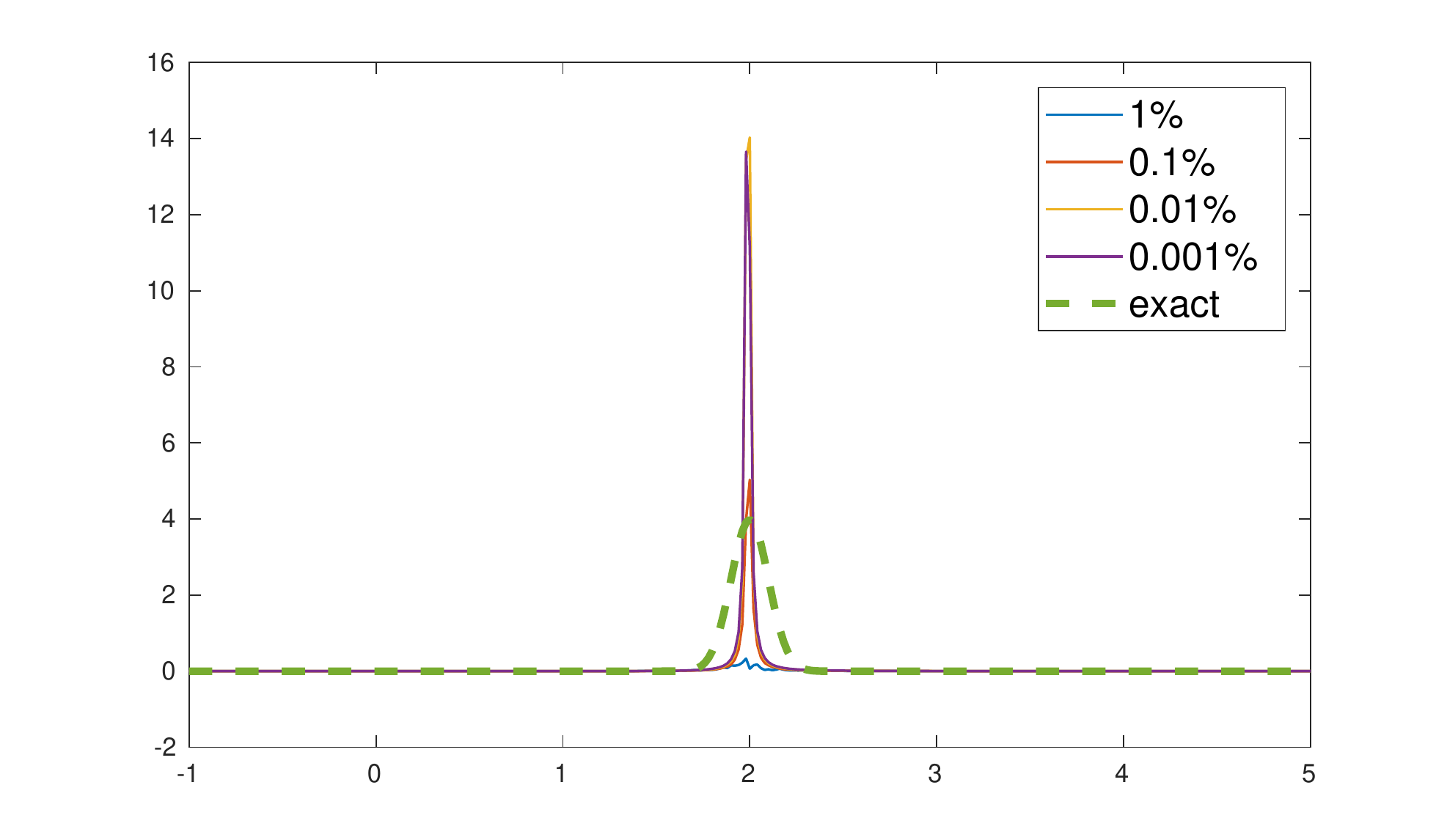}
	\caption{ Dashed line is the Gaussian distribution test spectrum. Colored solid lines are recovered spectra with different input errors.  As the input accuracy is increasing, neither $\text{mean}\{\rho(\omega)\}$ nor $\text{std}\{\rho(\omega)\}$ converge. Gaussian analytic function is a very special case. } 
\end{figure}

\begin{figure}[H]
	\centering\includegraphics[width=1\linewidth]{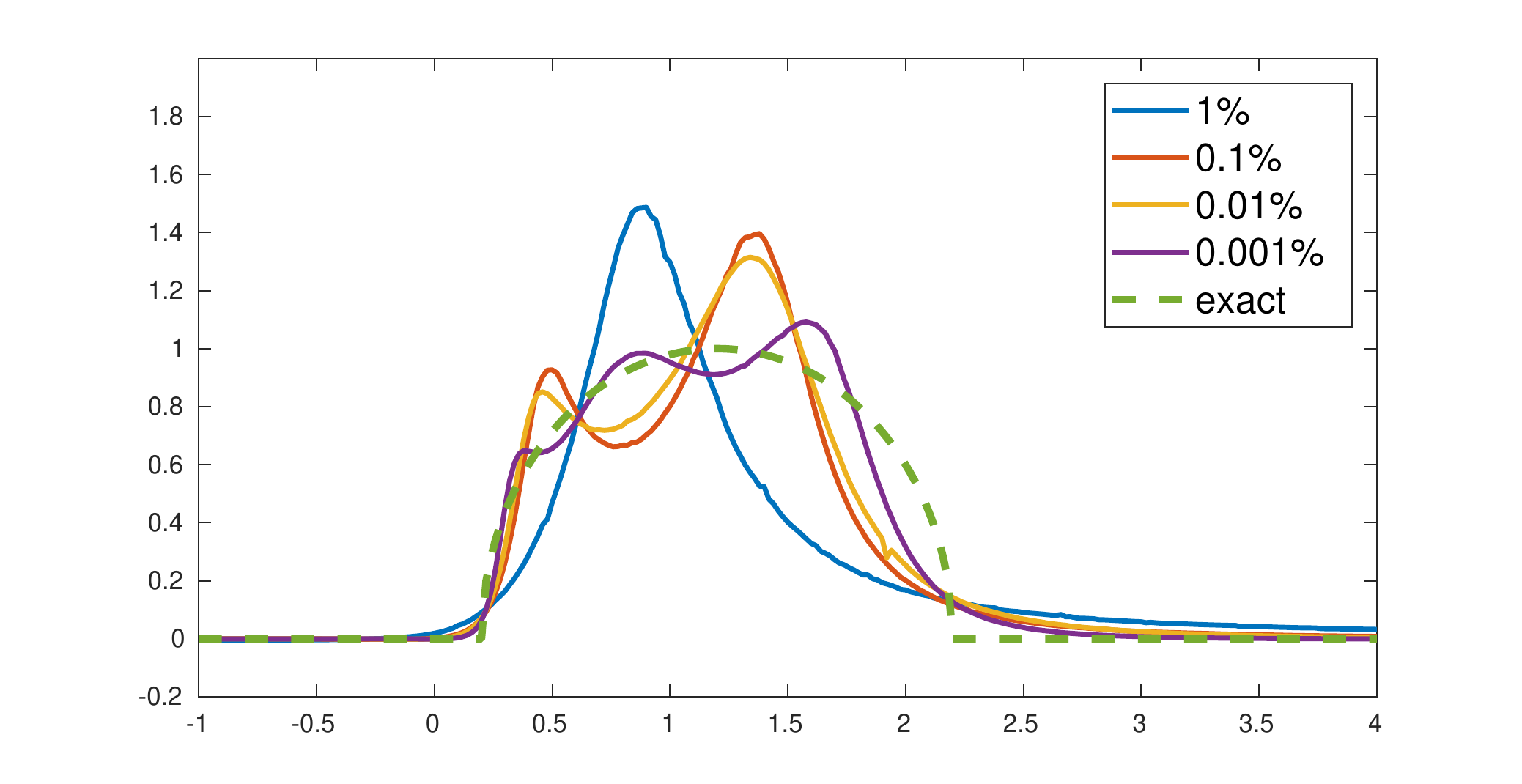}
	\caption{ Dashed line is the semicircle distribution test spectrum. Colored solid lines are recovered spectra with different input errors.} 
\end{figure}

\begin{figure}[H]
	\centering\includegraphics[width=1\linewidth]{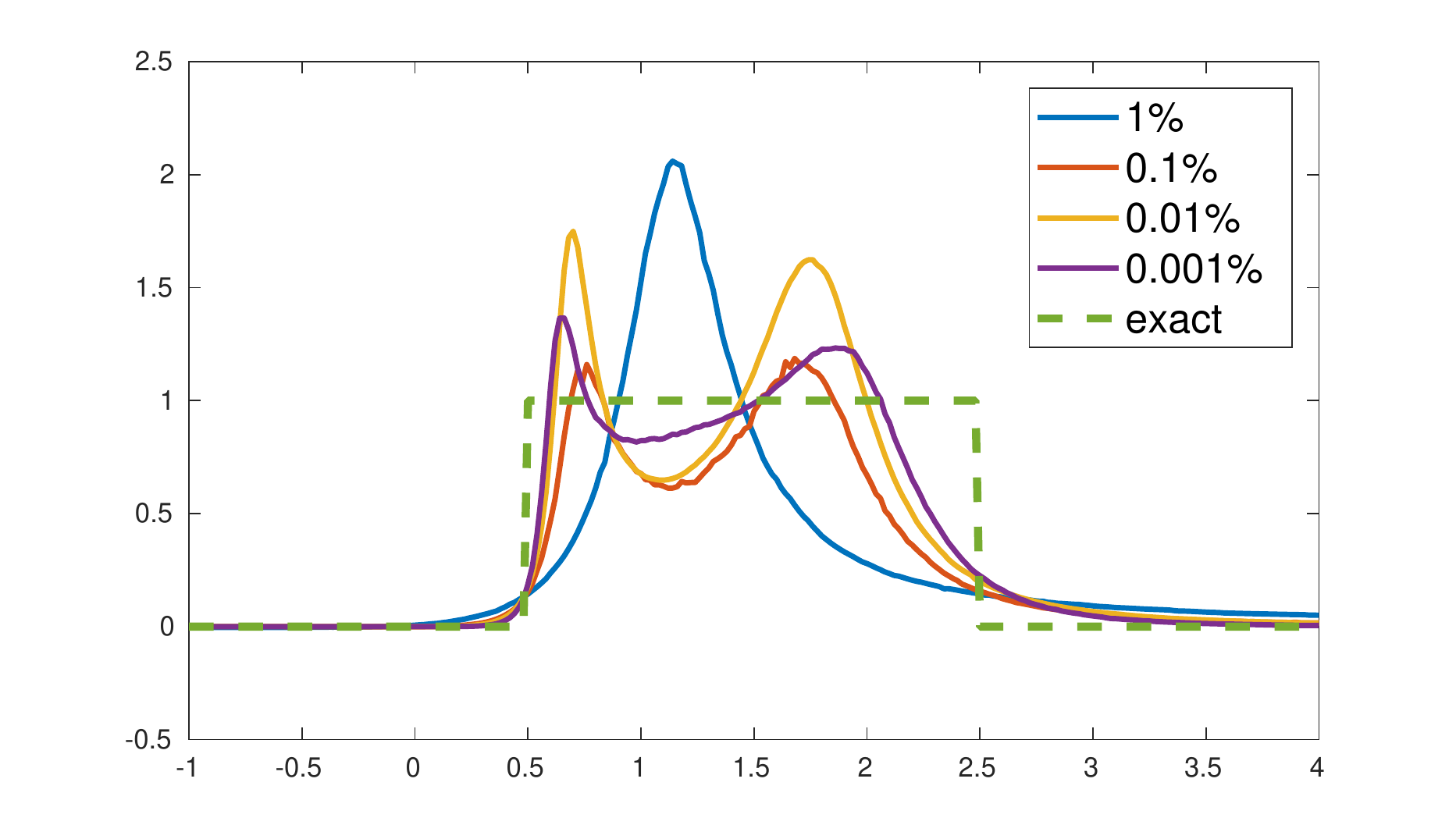}
	\caption{Dashed line is the square distribution test spectrum. Colored solid lines are recovered spectra with different input errors.} 
\end{figure}

\end{document}